\documentclass[
    ,final            
  ]
  {aipproc}

\layoutstyle{6x9}

\usepackage{graphicx}
\usepackage{subfigure}
\usepackage{epsfig}
\usepackage{graphics}
\usepackage{pslatex}

\newcommand{\sqrtsnn}{\sqrt{s_{_{\ensuremath{\it{NN}}}}}}
\def\mean#1{\ensuremath{\left<#1\right>}}

\begin{document}

\title{Jet quenching: RHIC results and phenomenology}



\author{David d'Enterria}{
  address={Nevis Laboratories, Columbia University\\ 
Irvington, NY 10533, and New York, NY 10027, USA}
}

\begin{abstract}
I review the main experimental results on jet physics in high-energy 
nucleus-nucleus collisions as studied via inclusive leading hadron 
spectra and di-hadron correlations at high transverse momentum. 
In central Au+Au at RHIC ($\sqrtsnn$ = 200 GeV), the observed large 
suppression of high-$p_T$ hadron spectra as well as the strongly 
modified azimuthal dijet correlations compared to baseline p+p results 
in free space, provide crucial information on the thermodynamical and 
transport properties of QCD matter.
\end{abstract}

\maketitle


\section{Introduction}

\label{sec:intro}
The research program of high-energy heavy-ion physics is centered on
the study of the collective properties of extended quark-gluon systems.
By colliding two heavy nuclei at relativistic energies one expects to form
a hot and dense deconfined medium whose collective (color) dynamics can be 
quantitatively described by  QCD thermodynamics calculations on the lattice~\cite{karsch}. 
In this context, the main goal of the RHIC experiments is the production and 
study under laboratory conditions of the Quark Gluon Plasma (QGP) predicted to be 
formed when strongly interacting matter attains energy densities above 
$\varepsilon\approx$ 1 GeV/fm$^3$.
The production of such an extremely hot and dense partonic system 
should manifest itself in a variety of experimental signatures~\cite{harris96}. 
One of the first proposed QGP ``smoking guns'' was ``jet quenching''~\cite{bjorken82} 
i.e. the disappearance of the collimated spray of hadrons resulting from the fragmentation 
of a hard scattered parton due to its ``absorption'' in the dense medium produced 
in the reaction. Extensive theoretical work on high-energy parton propagation in 
a QCD environment~\cite{gyulassy90,bdmps,glv,wiedemann} has shown that 
the main mechanism of attenuation is of radiative nature: the traversing 
parton loses energy mainly by multiple gluon emission (``gluonstrahlung'').
Such a medium-induced {\it non-Abelian} energy loss should result in 
several observable experimental consequences:
\begin{description}
\item (i) {\bf depleted} production of {\bf high $p_T$ leading hadrons} ($dN/dp_T$)~\cite{gyulassy90}, 
\item(ii) {\bf unbalanced} back-to-back {\bf di-jet azimuthal correlations} 
($dN_{pair}/d\phi$)~\cite{appel86_blaizot_mclerran86}, and
\item(iii) {\bf modified energy flow} and {\bf particle multiplicity} within the 
final jets~\cite{armesto-salgado-wiedemann,borghini05}.
\end{description}
By quantitatively comparing the jet structure modifications in A+A relative to
baseline p+p collisions in free space, one can have experimental access to the
properties of the produced QCD matter. In this overview, we discuss several
significant experimental measurements from Au+Au reactions at RHIC which 
have been phenomenologically linked to key thermodynamical and transport properties 
which can, in some cases, be directly computed in lattice QCD.
E.g., if the observed high-$p_T$ leading hadron suppression is due to 
medium-induced gluon radiation off hard scattered partons, then
\begin{itemize}
\item the initial {\bf gluon density} $dN^g/dy$ of the expanding plasma (with transverse
area $A_\perp$ and length $L$) can be estimated from the measured energy loss via~\cite{glv}:
\begin{equation}
\Delta E \propto \alpha_S^3\,C_R\,\frac{1}{A_\perp}\frac{dN^g}{dy}\,L\mbox{ ,}
\label{eq:glv}
\end{equation}
\item and also the {\bf transport coefficient} $\langle\hat{q}\rangle$,
characterizing the squared average momentum transfer from the medium to the hard parton
per unit distance, can be derived from the average energy loss according to~\cite{bdmps,wiedemann}:
\begin{equation}
\langle\Delta E\rangle \propto \alpha_S\,C_R\,\langle\hat{q}\rangle\,L^2.
\label{eq:bdmps}
\end{equation}
\end{itemize}
Likewise, it has been argued that a fast parton propagating (and loosing energy) 
through the medium can generate a wake of lower energy gluons with Mach-~\cite{mach,rupp05}
or \v{C}erenkov-like~\cite{rupp05,cerenkov} conical angular patterns.
\begin{itemize}
\item In the first case, the {\bf speed of sound} of the traversed matter, 
$c_s^2 = \partial P/\partial\varepsilon$, can be determined from the characteristic angle 
of the emitted secondaries with respect to the (quenched) jet axis~\cite{mach,rupp05}:
\begin{equation}
\cos(\theta_{M}) = c_s\;\mbox{ , where $\theta_{M}$ is the Mach shock wave angle}.
\label{eq:mach}
\end{equation}
\item In the second scenario, the {\bf gluon dielectric constant} in the medium $\varepsilon$ 
or, equivalently, its index of refraction $n = \sqrt{\varepsilon}$, can be estimated 
from the distinctive angular pattern of emission of the (soft) radiated gluons~\cite{rupp05,cerenkov}:
\begin{equation}
\cos(\theta_{c})\approx \frac{1}{\sqrt{\epsilon}} \approx \frac{1}{n}
\label{eq:cerenkov}
\end{equation}
\end{itemize}

\section{High $p_T$ leading hadron suppression}

\begin{figure}[htbp]
\includegraphics[width=8.0cm]{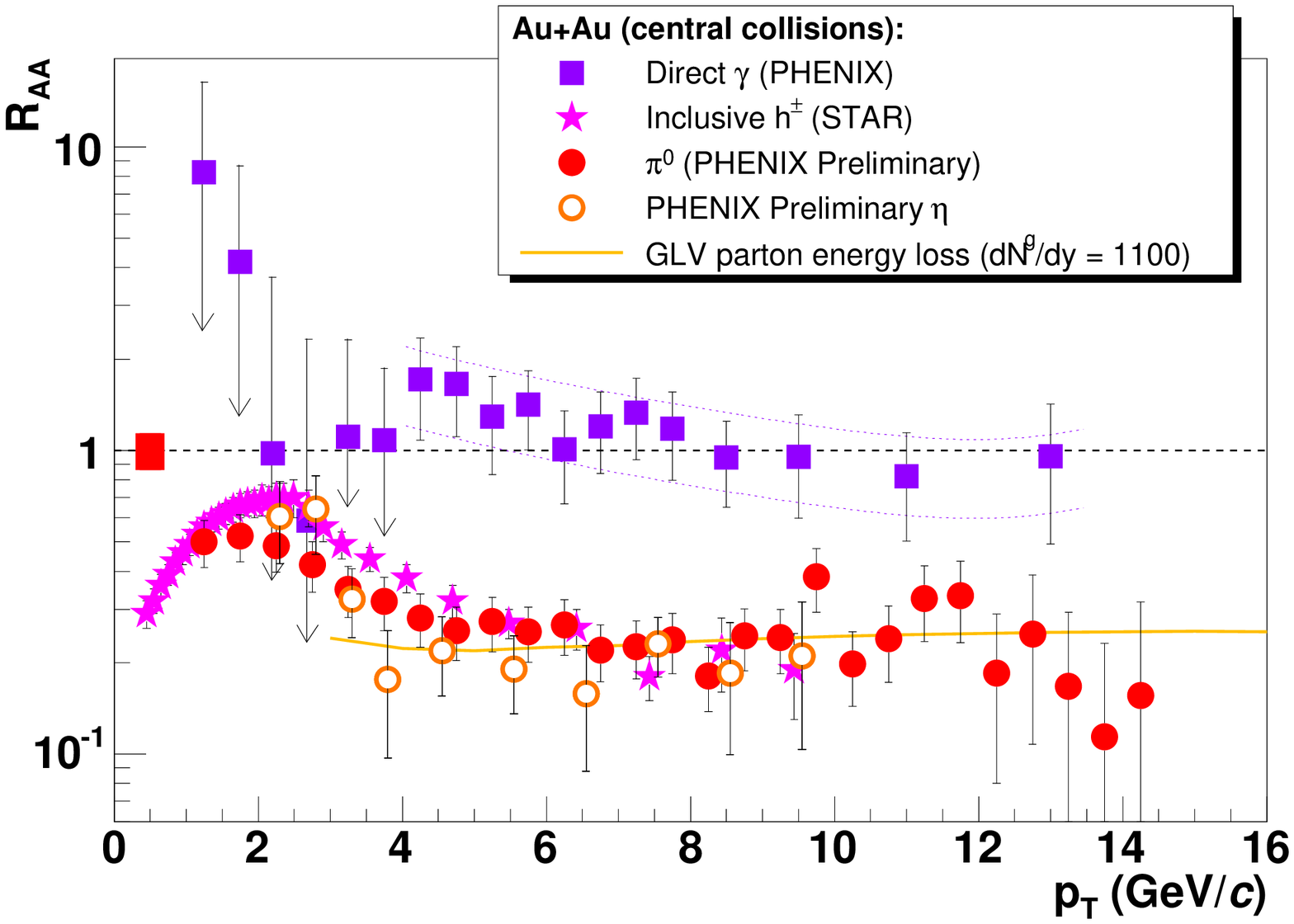}
\includegraphics[width=8.5cm]{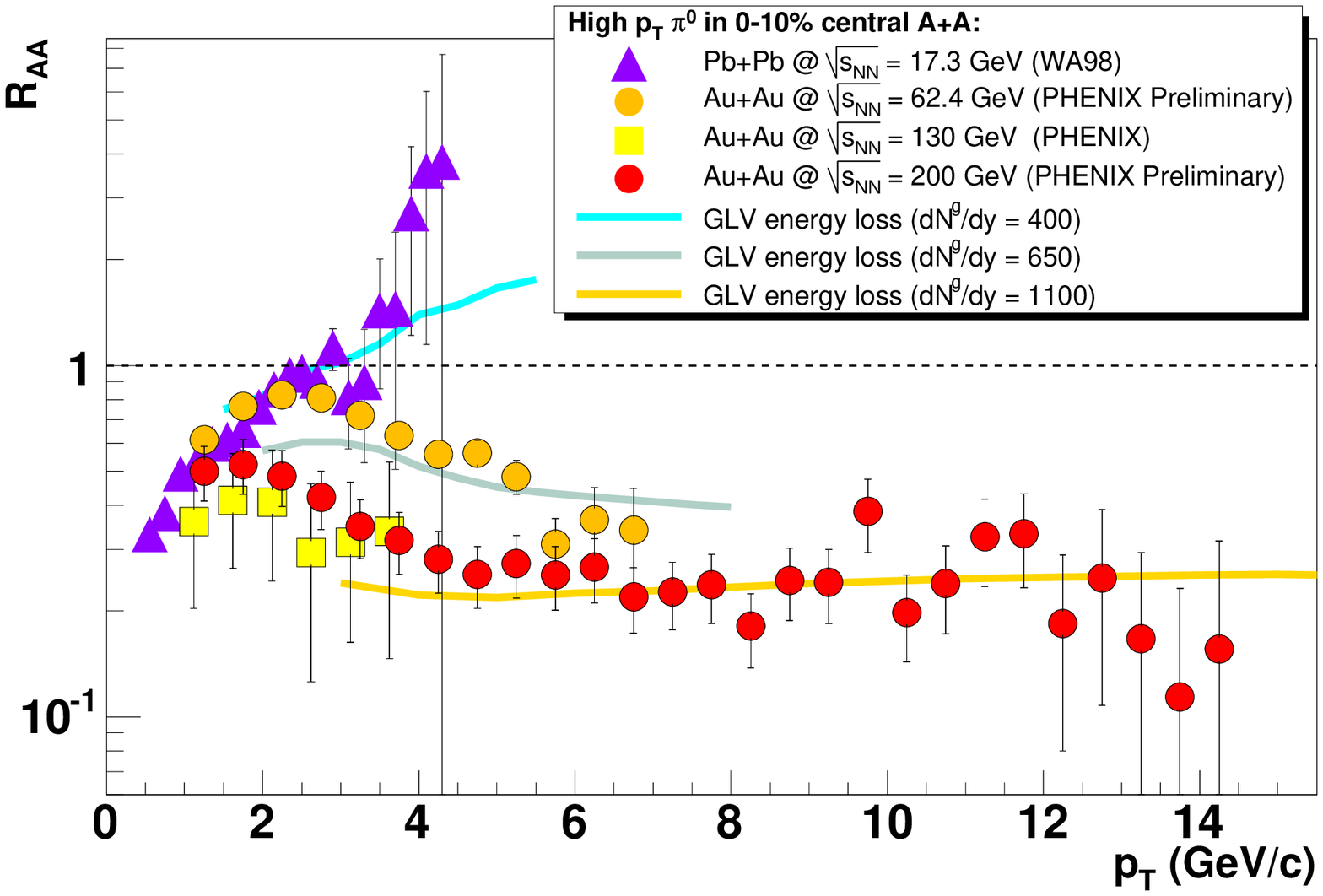}
\caption{Left: $R_{AA}(p_T)$ measured in central Au+Au at 200 GeV for: 
$\pi^0$ and $\eta$ mesons~\protect\cite{phenix_hipt_pi0_eta_200},
charged hadrons~\protect\cite{star_hipt_200},
and direct photons~\protect\cite{phenix_AuAugamma} 
compared to theoretical predictions for parton energy loss in a dense medium with 
$dN^g/dy=$ 1100~\protect\cite{vitev_gyulassy}.
Right: Compilation of all measured $R_{AA}(p_T)$ for high $p_T$ neutral pions 
in central A+A collisions in the range $\sqrtsnn\approx$ 20 -- 200 GeV~\protect\cite{dde_hp04},
compared to GLV parton energy loss calculations~\protect\cite{vitev_gyulassy} 
for different initial gluon densities ($dN^g/dy$ = 400, 650 and 1100).}
\label{fig:R_AA_RHIC_200}
\end{figure}

\begin{figure}[htbp]
\includegraphics[width=8.5cm]{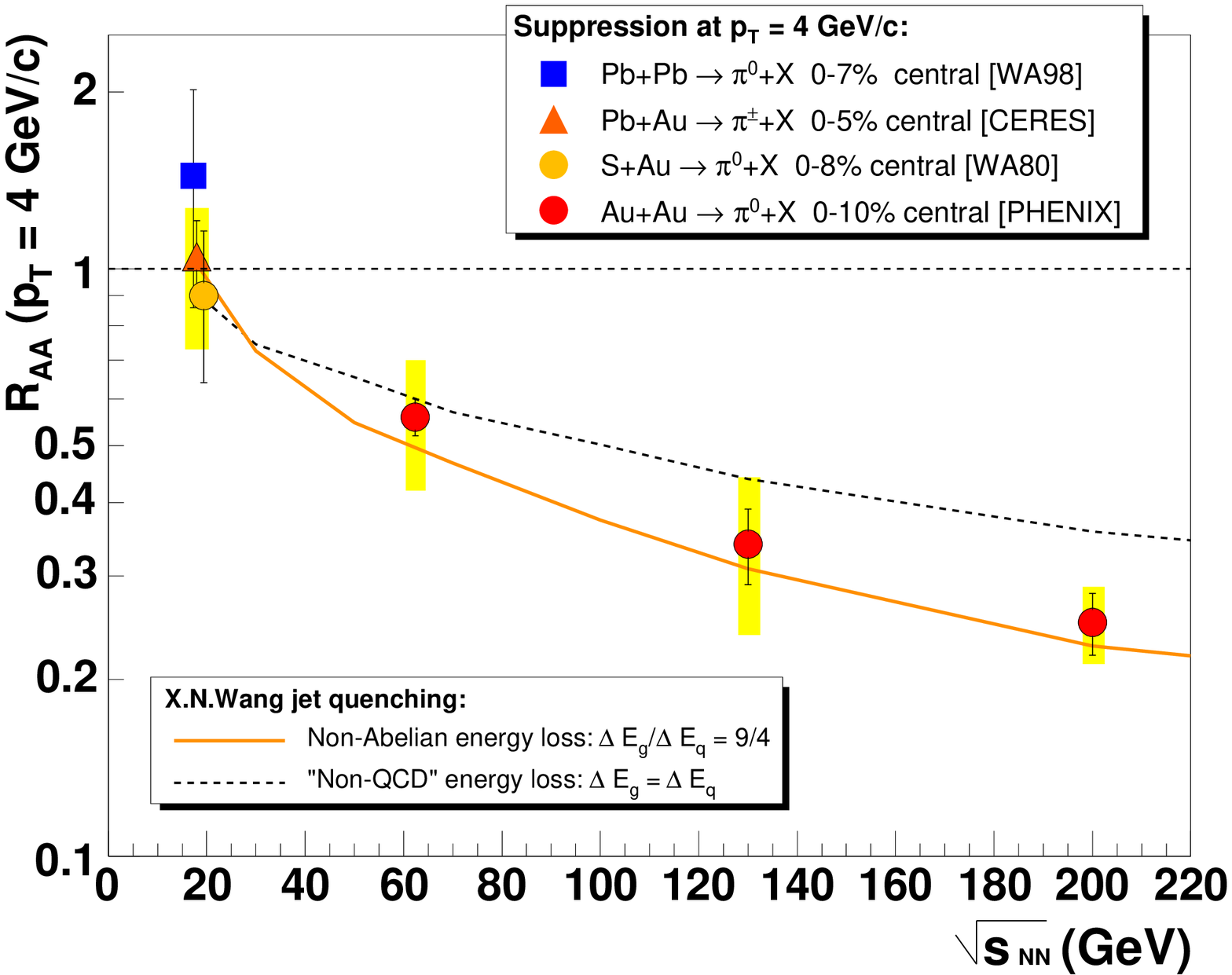}
\includegraphics[width=8.0cm,height=6.5cm]{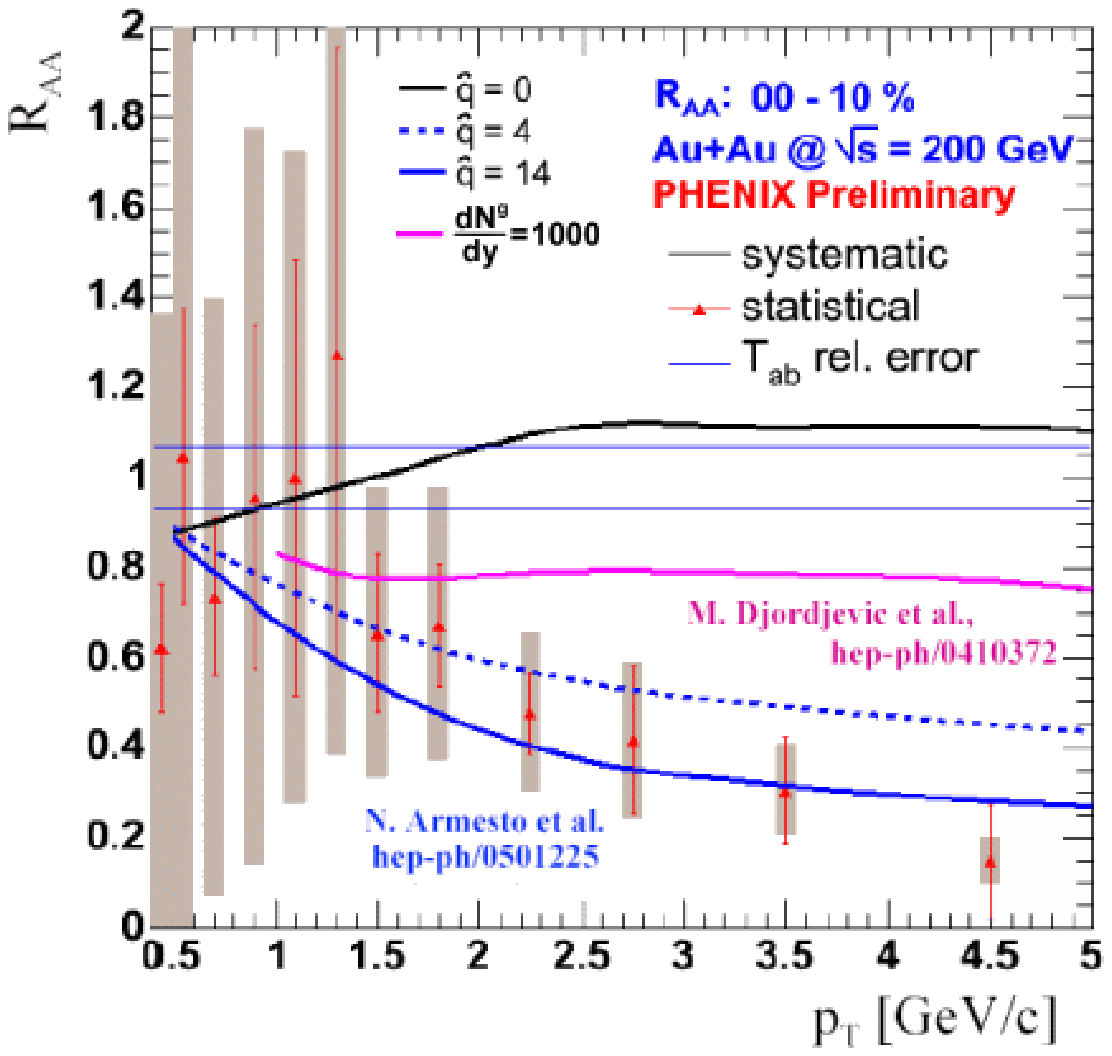}
\caption{Left: Excitation function of $R_{AA}(p_T = 4$ GeV/$c)$ for $\pi^0$~\protect\cite{dde_hp04} 
with two different implementations of partonic energy loss~\protect\cite{wang05}: 
(i) canonical non-Abelian (gluons loose $C_A/C_F=9/4$ more energy than quarks; solid line) 
and (ii) {\it ad hoc} ``non-QCD'' ($q,g$ radiate with equal probability; dashed line) prescriptions.
Right: $R_{AA}$ for ``non-photonic'' $e^\pm$ measured in central Au+Au at $\sqrtsnn$ = 200 GeV~\protect\cite{jacak05}
compared to theoretical predictions of heavy-quark energy loss~\protect\cite{djordj04,armesto05}.}
\label{fig:RAA_vs_sqrts}
\end{figure}

The standard method to quantify the (initial- and final-state) medium effects 
on the production yields 
of a given hard probe in a nucleus-nucleus reaction is given by the 
{\it nuclear modification factor}:
\begin{equation} 
R_{AA}(p_{T},y;b)\,=\,\frac{\mbox{\small{``hot/dense QCD medium''}}}{\mbox{\small{``QCD vacuum''}}}\,= \,
\frac{d^2N_{AA}/dy dp_{T}}{\langle T_{AA}(b)\rangle\,\cdot\, d^2 \sigma_{pp}/dy dp_{T}},
\label{eq:R_AA}
\end{equation}
which measures the deviation of A+A at impact parameter $b$ from an incoherent 
superposition of nucleon-nucleon collisions ($T_{AA}(b)$ is the corresponding Glauber nuclear 
overlap function at $b$). Among the most exciting results from the first 5 years of 
operation at RHIC is the large high $p_T$ hadron suppression ($R_{AA}\ll$ 1) observed in central Au+Au 
reactions at $\sqrtsnn$ = 200 GeV, expected in jet quenching scenarios. Most of the 
empirical properties of the suppression factor are in quantitative agreement with the 
predictions of non-Abelian parton energy loss models: 

\begin{description}
\item (1) {\bf Magnitude} of the suppression: The experimental $R_{AA}\approx$ 0.2 
value at top RHIC energies can be well reproduced assuming the formation of a 
very dense system with initial gluon rapidity density
$dN^g/dy\approx$ 1000~\cite{vitev_gyulassy} (Eq.~\ref{eq:glv}) or transport 
coefficient $\mean{\hat{q}}\approx$ 14 GeV$^2$/fm~\cite{dainese04} (Eq.~\ref{eq:bdmps}), 
both consistent with the total charged hadron multiplicities measured in the reaction:
$dN/d\eta\approx 3/2\cdot dN_{ch}/d\eta\approx$ 1000~\cite{dde_hp04}.

\item (2) {\bf Universal} (light) hadron suppression: Above $p_T\approx$ 5 GeV/$c$, 
$\pi^0$~\cite{phenix_hipt_pi0_200}, $\eta$~\cite{phenix_hipt_pi0_eta_200}, 
and inclusive charged hadrons~\cite{star_hipt_200,phenix_hipt_200} 
(dominated by $\pi^\pm$~\cite{phenix_hipt_200}) show all a common factor 
of $\sim$5 suppression relative to the $R_{AA}$ = 1 perturbative expectation
which holds for hard probes, such as direct photons, insensitive to 
final-state interactions~\cite{phenix_AuAugamma} (Fig.~\ref{fig:R_AA_RHIC_200}, left).
Such a ``universal'' hadron deficit is consistent with in-medium {\it partonic} 
energy loss of the parent quark or gluon prior to vacuum fragmentation.

\item (3) Flat {\bf transverse momentum} dependence: 
Above $p_T\approx$ 5 GeV/$c$, $R_{AA}(p_T)$ remains constant up to the highest 
transverse momenta measured so far ($p_T\approx$ 14 GeV/$c$ for $\pi^0$~\cite{phenix_hipt_pi0_eta_200}, 
Fig.~\ref{fig:R_AA_RHIC_200} left). Such $p_T$-independence of the quenching factor
is also well accommodated by parton energy loss models~\cite{vitev_gyulassy,dainese04,jeon_moore}.

\item (4) {\bf Centrality} dependence: The amount of suppression in Au+Au reactions
decreases with impact parameter as expected (from Eqs.~\ref{eq:glv},\ref{eq:bdmps}) 
for the different parton production points and, hence, the different densities and 
lengths encountered by the traversing parton through the medium~\cite{dainese04,wang04}.

\item (5) {\bf Center-of-mass energy} dependence: The amount of quenching rises 
in the range $\sqrtsnn\approx$ 20 -- 200 GeV as expected due to the growing initial 
parton densities and the increasingly longer duration of the QGP phase~\cite{dde_hp04} 
(Fig.~\ref{fig:R_AA_RHIC_200}, right).

\item (6) {\bf Non-Abelian nature} of the energy loss: At $y=0$, high-$p_T$ hadroproduction 
is dominated by quark (gluon) scattering at large (small) fractional momentum $x_T = 2p_T/\sqrtsnn$. 
In the range $\sqrtsnn\approx$ 20 -- 200 GeV and for a {\it fixed} (high) $p_T$ value, 
the suppression factor increases as expected in the canonical non-Abelian scenario where 
there is an increasingly large relative fraction of hard scattered gluons radiating with 
a $C_A/C_F=9/4$ larger probability than quarks (Fig.~\ref{fig:RAA_vs_sqrts}, left).

\end{description}

Though most of the experimental results on high $p_T$ hadron suppression in A+A reactions 
are in agreement with the 
non-Abelian energy loss ``paradigm'', it is worth to stress that there are a 
few aspects of the data which are less well reproduced:

\begin{description}
\item (1) The pronounced {\bf path-length} $L$ dependence of the energy loss, 
-- as determined by the $\pi^0$ suppression factor along different azimuthal 
angles $\phi$ with respect to the Au+Au reaction plane~\cite{cole} --, does 
not support the theoretical $\propto L$ or $L^{2}$ behaviour given by 
Eqs.~\ref{eq:glv}, \ref{eq:bdmps}. Such a failure of the jet-quenching 
models points to an extra source of azimuthal anisotropy that enhances the in-plane 
$\pi^0$ production even in a kinematic domain beyond $p_T\approx$ 4.5 GeV/$c$
where non-perturbative effects should play a minor role~\cite{dde_hp04}.

\item (2) The {\bf unsuppressed baryon} ($p,\bar{p}$~\cite{phenix_hipt_ppbar} and 
$\Lambda,\bar{\Lambda}$~\cite{star_hipt_lambdas}) spectra within $p_T\approx$ 2 -- 5 GeV/$c$ 
have been explained in terms of an extra mechanism of baryon production 
based on in-medium quark coalescence~\cite{reco}
which compensates for the energy loss suffered by the fragmenting parent partons.
Though such a mechanism successfully describes many aspects of the data,
it cannot explain (in its simplest ``thermal recombination'' form) 
the similar (``jet-like'') azimuthally-correlated hadron yields measured for 
trigger baryons and mesons at intermediate $p_T$~\cite{phenix_jet_baryons}.

\item (3) The {\bf large} amount of {\bf heavy-quark quenching} indicated by the suppressed 
high-$p_T$ spectra of electrons from semi-leptonic $D$ and $B$ meson decays 
measured by PHENIX in central Au+Au~\cite{jacak05} (Fig.~\ref{fig:RAA_vs_sqrts}, right) 
is in apparent conflict with the robust $\Delta E_{Q} < \Delta E_{q} <  \Delta E_{g}$ 
prediction of parton energy loss models. State-of-the-art theoretical 
predictions~\cite{djordj04,armesto05} require much larger gluon densities to 
reproduce the high $p_T$ open charm/bottom results at RHIC, than they needed to 
describe the quenched light hadron spectra.

\end{description}

\section{Modified high $p_T$ di-hadron $\phi,\eta$ correlations}

\begin{figure}[htbp]
\includegraphics[width=7.5cm,height=7.cm]{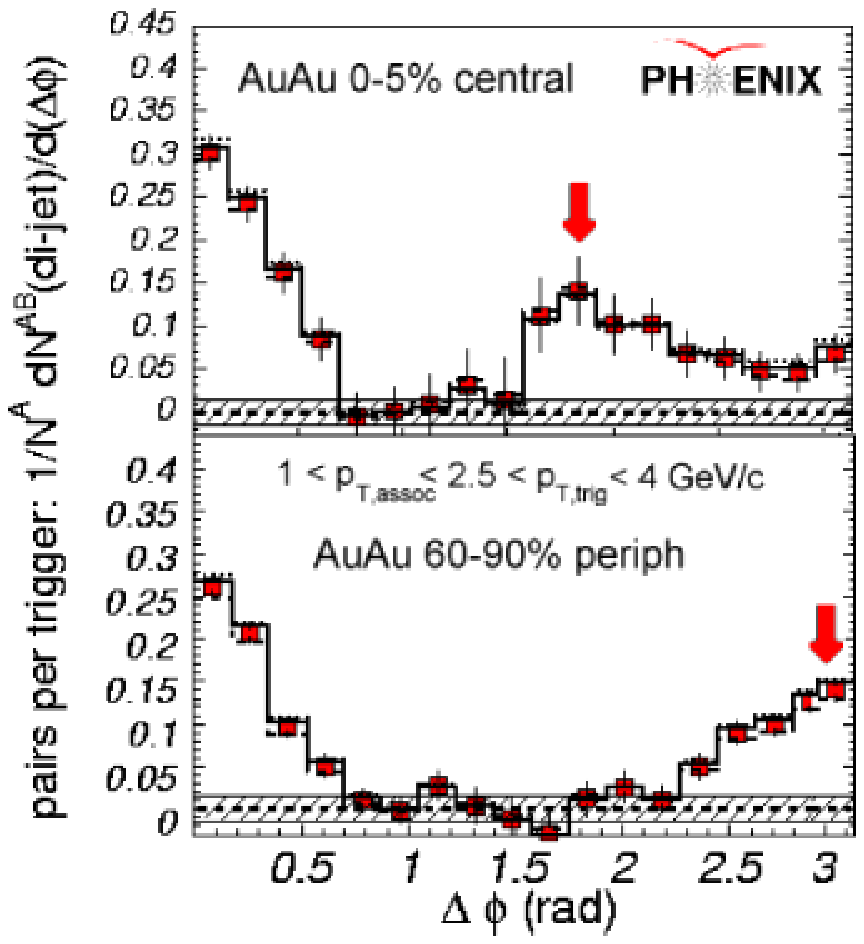}
\includegraphics[width=7.5cm,height=7.cm]{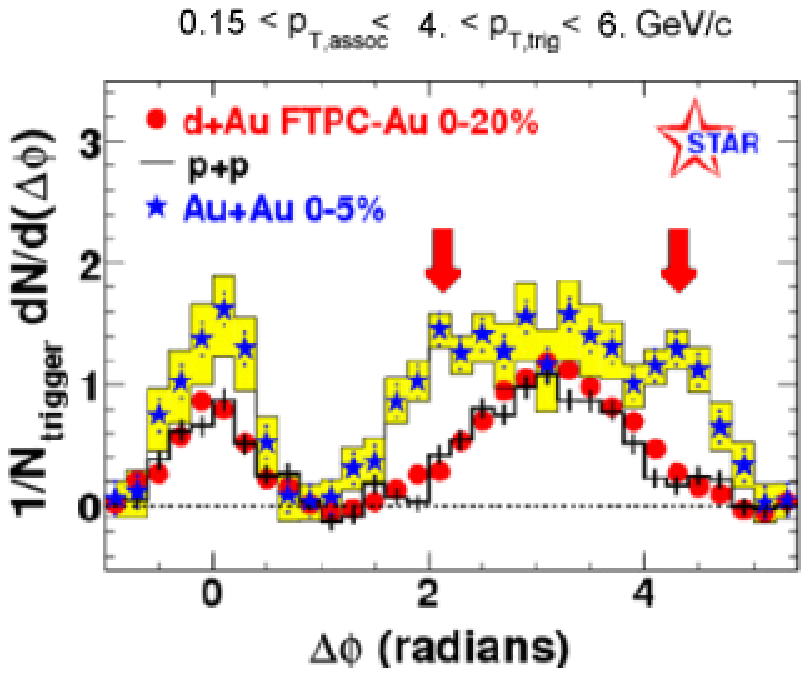}
\caption{Azimuthal dihadron distributions normalized per trigger particle, 
$1/N_{trigg}\,dN_{pair}/d\Delta\phi$ (the arrows indicate the local maxima in the 
away-side hemisphere) measured at RHIC. Left: PHENIX results in central (top) 
and peripheral (bottom) Au+Au~\protect\cite{phenix_machcone}. 
Right: STAR results in central Au+Au, d+Au and p+p collisions~\protect\cite{star_hipt_etaphi}.
(Note that each experiment has different $p_T$ values for the associated hadrons).}
\label{fig:dNdphi}
\end{figure}

Full jet reconstruction in A+A collisions with standard jet algorithms~\cite{jet_algo} 
is unpractical at RHIC energies due to the overwhelming background 
of soft particles in the underlying event. Thus, beyond the leading hadron spectra 
measurements discussed in the previous section, more detailed studies of the
modifications of the jet structure in a dense QCD environment have been addressed 
via the study of high-$p_T$ two-particle $\phi,\eta$ correlations. 
Jet-like correlations are measured on a statistical basis by selecting high-$p_T$ 
{\it trigger} particles and measuring the azimuthal ($\Delta\phi = \phi - \phi_{trig}$)
and rapidity ($\Delta\eta = \eta - \eta_{trig}$) distributions of 
{\it associated} hadrons ($p_{T,assoc}<p_{T,trig}$) relative to the trigger:
\begin{equation}
C(\Delta\phi,\Delta\eta) = \frac{1}{N_{trig}}\frac{d^2N_{pair}}{d\Delta\phi d\Delta\eta}.
\end{equation}
Combinatorial background contributions, corrections for finite pair acceptance, and
the superimposed effects of {\it collective} azimuthal modulations (elliptic flow)
are then taken care of with different techniques~\cite{star_hipt_awayside,star_hipt_etaphi,phenix_machcone}. 
If no initial- or final- state interactions affect the parton-parton scattering process, 
a dijet signal should appear to first order as two distinct back-to-back Gaussian 
peaks at $\Delta\phi\approx$ 0, $\Delta\eta\approx$ 0 (near-side) and at 
$\Delta\phi\approx\pi$ (away-side). At variance with this standard dijet topology
in the QCD vacuum, the di-hadron correlations in Au+Au reactions at RHIC show several 
striking features:
\begin{description}
\item (1) The gradual {\bf disappearance} of the {\bf away-side azimuthal peak} with centrality 
(observed at $\Delta\phi\approx\pi$ in the $dN_{pair}/d\Delta\phi$ distributions
for hadrons with $2<p_{T,assoc}<4 < p_{T,trig} < 6$ GeV/$c$), consistent with strong 
suppression of the leading fragments of the recoiling jet traversing the medium~\cite{star_hipt_awayside}.
\item (2) The {\bf broadening} of the {\bf nearside pseudo-rapidity} correlations $dN_{pair}/d\Delta\eta$ 
(``stretching'' of the jet cone along $\eta$), reminiscent of the coupling 
of the induced radiation with the longitudinal expansion of the system~\cite{star_hipt_etaphi}.
\item (3) The vanishing away-side peak, observed in the $dN_{pair}/d\Delta\phi$ 
distribution for recoiling hadrons with $p_{T,assoc}$ = 2 -- 4 GeV/$c$, is accompanied with an 
{\it enhanced} production of {\it lower} $p_T$ hadrons ($p_{T,assoc}$ = 1 -- 2.5 GeV/$c$~\cite{phenix_machcone}
or 0.15 -- 4 GeV/$c$~\cite{star_hipt_etaphi}) with a characteristic {\bf ``double-peak'' structure} 
at $\Delta\phi\approx\pi\pm$ 1.3 or $\pi\pm$ 1.1 (Fig.~\ref{fig:dNdphi}).
\end{description}

Figure~\ref{fig:dNdphi} shows the double-peak structure appearing in the 
away-side azimuthal correlations of central Au+Au (top left-plot, 
and star symbols in the right-plot), compared to the standard back-to-back dijet 
topology seen in peripheral Au+Au (bottom, left) and in d+Au and p+p collisions (right).
Such a non-Gaussian ``volcano''-like shape seen in the away-side hemisphere has attracted 
much theoretical attention because it suggests 
conical patterns induced by Mach-shock~\cite{mach,rupp05} or \v{C}erenkov-like~\cite{rupp05,cerenkov} 
emissions.

In the ``Mach cone'' scenario~\cite{mach,rupp05}, the local maxima (red arrows in the plots) 
in central Au+Au at an angle $\Delta\phi \approx \pi\pm 1.2$ relative to the high-$p_T$ 
trigger are caused by the Mach shock of the supersonic recoiling (quenched) parton through 
the medium. The resulting preferential emission of secondary partons from the plasma at an 
angle $\theta_{M} \approx 1.2$, 
yields (Eq.~\ref{eq:mach}) a value of the speed sound $c_{s}\approx$ 0.36, close to that 
of an ideal QGP ($c_s = 1/\sqrt{3}$). In the \v{C}erenkov gluonstrahlung picture~\cite{rupp05}, 
-- developed at a more quantitative level~\cite{cerenkov} after this conference --, 
it is argued that the combination of the 

\noindent
Landau-Pomeranchuk-Migdal interference characteristic of gluon bremsstrahlung and 
a medium with a {\it large} dielectric constant ($n\approx$ 2.75 is needed to 
reproduce the location of the experimental peaks using Eq.~\ref{eq:cerenkov}) 
should also result in the two-peak shape observed in the data. At variance with the 
cone angle of the ``sonic boom'' mechanism (which is constant in the fluid but 
effectively {\it increases} with $p_{T,assoc}$ at the spectra level~\cite{mach}), 
the \v{C}erenkov angle {\it decreases} with the momentum of the radiated gluon. 
Such a trend is, however, seemingly in disagreement with the fact that PHENIX (STAR)
measures a {\it larger} (lower) $\theta_{c}\approx$ 1.3 (1.1) for higher (lower) 
average values of $p_{T,assoc}$.



\section{Summary}

Experimental results on single inclusive spectra and dihadron correlations 
measured at high transverse momentum in Au+Au at RHIC collider energies 
($\sqrtsnn$ = 200 GeV) have been reviewed 
as a means to learn about jet production and fragmentation in hot and dense QCD matter. 
The analysis of jet structure modifications in A+A collisions provides 
quantitative information on the thermodynamical and transport properties of the 
strongly interacting medium produced in the reactions.
Two notable experimental results have been discussed: 
(i) the observed factor $\sim$5 suppression of high $p_T$ leading hadrons 
in central Au+Au relative to p+p collisions in free space; and 
(ii) the conical-like shape of the azimuthal distributions of secondary hadrons 
emitted in the away-side hemisphere of a high-$p_T$ trigger hadron.
Most of the properties of the observed high $p_T$ suppression (such as its magnitude,
light flavor ``universality'', $p_T$, reaction centrality, and $\sqrtsnn$ dependences) 
are in quantitative agreement with predictions of non-Abelian energy loss models.
The confrontation of these models to the data permits to derive the initial 
gluon density $dN^g/dy\approx$ 1000 and transport coefficient 
$\mean{\hat{q}}\approx$ 14 GeV$^2$/fm of the produced medium.
The second striking observation of a softer and broadened angular distribution of 
secondary hadrons peaking at a finite angle away from the (quenched) jet axis 
has been attributed to Mach conical flow caused by the propagation of 
a supersonic parton through the dense system. If such a phenomenon is confirmed, 
the speed of sound of the medium could be extracted. The same angular pattern 
could also be the result of \v{C}erenkov gluon radiation and provide, in that case, 
information on the gluon dielectric constant in hot and dense QCD matter.


\end{document}